\documentclass[preprint,12pt]{elsarticle}
\usepackage{amssymb}
\newcommand{\ba}{\begin{eqnarray}}
\newcommand{\ea}{\end{eqnarray}}
\newcommand{\beqs}{\begin{eqnarray}}
\newcommand{\eeqs}{\end{eqnarray}}

\begin{document}

\begin{frontmatter}

\title{ Electromagnetic and gravitomagnetic structure and radii of nucleons  \footnote{talk on The XXIV International Workshop “High Energy Physics and Quantum Field Theory” (QFTHEP 2019)}  } %  \jpcs}

%\author{Oleg Selyugin}

%\jpcs, \iopp,

\author{ O.V. Selyugin   \footnote{%talk on The XXIV International Workshop “High Energy Physics and Quantum Field Theory” (QFTHEP 2019) \\ 
e-mail:  selugin@theor.jinr.ru} } % \thanks{selugin@theor.jinr.ru} }
%\thanks{\email{selugin@theor.jinr.ru}} }  % \inst{1,3}   % \fnsep
%\institute{BLTPh, JINR, Dubna, Russia}

\address{ BLTPh, JINR,  Dubna, Russia}

%\ead{selugin@theor.jinr.ru}

\begin{abstract}
%\abstract{
Taking into account the PDFs, obtained by different Collaborations,
  the momentum transfer dependence of GPDs of the nucleons % and mesons
  is obtained.
   The calculated   electromagnetic and gravitomagnetic
  form factors of  nucleons are used for the description of  different form factors and
  the  nucleons % hadron-hadron
   elastic scattering in a wide energy and momentum transfer region
  with a minimum number of fitting parameters.
%The electromagnetic and gravitomagnetic densities of the quarks
% in nucleons and mesons   are calculated.
  The electromagnetic and gravitomagnetic radii of the nucleon are calculated using the obtained
  momentum transfer dependence of GPDs with different forms of PDFs obtained by  different
  Collaborations.
  The comparison of the calculations,
  taking into account the PDFs obtained  by different Collaborations,
  of mean square electromagnetic and gravitomagnetic  radii  of nucleons % are made.
  % and meson the momentum transfer dependence of GPDs of the pion
   is made. % obtained.

%Different models of the hadron structure are analyzed.
%  Taking into account the PDFs, obtained by different Collaborations,
%  the momentum transfer dependence of GPDs of the nucleons and mesons
%  is obtained. The electromagnetic and gravitomagnetic densities of the quarks
% in nucleons and mesons   are calculated. The comparison of the calculations,
%  taking into account the PDFs obtained  by different Collaborations,
%  of mean square electromagnetic and gravitomagnetic  radii  of nucleons % are made.
%  % and meson the momentum transfer dependence of GPDs of the pion
%   is make. % obtained.
%   The calculated   electromagnetic and gravitomagnetic
%  form factors of the nucleons are used for the description of  different form factors and
%  the  hadron-hadron   elastic scattering in a wide energy and momentum transfer region
%  with a minimum number of fitting parameters.
%  The electromagnetic and gravitomagnetic radii of the nucleon are calculated using the obtained
%  momentum transfer dependence of GPDs with different forms of PDFs obtained by  different
%  Collaborations.
  %}
\end{abstract}

%    The calculated   electromagnetic and gravitomagnetic
%  form factors of the  pions and nucleons are used for the description
%  of the  pion-nucleon   elastic scattering in wide energy and momentum transfer region
%  with minimum fitting parameters.}
%\end{abstract}

%\maketitle
\begin{keyword}
    high energies, hadrons, radii, electromagnetic structure, matter distribution
% MSC codes here, in the form: \MSC code \sep code
% or \MSC[2008] code \sep code (2000 is the default)
\end{keyword}

% \end{document}
\end{frontmatter}
%\end{document}

\section{Introduction}

 One of the most important tasks of the Standard model is the research of the structure of "elementary" particles.
 The main integral properties of  particles are  reflected in their radii.
 This is tightly connected with the particle form factors.
% For the hadrons there are the Dirac and Pauly form factors
% which are determined as
 %  In the first moment, it is connected with the electromagnetic structure of the nucleon
   This is primarily due to the electromagnetic structure of the nucleon
   which can be obtained from the electron-hadron elastic scattering.
   In the Born approximation,
       the Feynman amplitude for the elastic electron-proton scattering \cite{Lomon} is
  \ba
M_{ep\rightarrow ep} = \frac{1}{q^2}[e \bar{u}(k_2)\gamma^{\mu} u(k_{1}][e\bar{U}(p_{2}\Gamma_{\mu}(p_{1},p_{2})U(p_{1}],
 \ea
   where $u$ and $U$ are the electron and nucleon  Dirac spinors,
\ba
\Gamma^{\mu} = F_{1}(t)\gamma^{\mu} \ + \  F_{2}(t) \frac{i\sigma^{\mu\nu} q_{\nu}}{2 m},
\ea
   where $m$ is the nucleon mass, $\kappa$ is the anomalous part
   of the magnetic moment and $t=-q^2 =-(p-p^{\prime})^2$ is the square of the momentum transfer of the nucleon.
   The functions $F_{1}(t)$ and $F_{2}(t)$ are named  the Dirac and Pauli form factors, which depend upon the nucleon structure.
   The normalization of the form factors  \cite{QA12}  is given by
   \ba
   F_{1}^{p}(t=0)=1, \ \ \ F_{2}^{p}(t=0)=\kappa_{p}=1.793
   \ea
   for the proton and
   \ba
   F_{1}^{n}(t=0)=0, \ \ \ F_{2}^{n}(t=0)=\kappa_{n}=-1.913
   \ea
   for the neutron.

 Two important combinations of the  Dirac and Pauli form factors
are the so-called Sachs form factors \cite{Ernst60,Sachs62}.
   In the Breit frame the current is separated into the
   electric and magnetic contributions \cite{Kelly-02}
   \ba
  \bar{u}(p^{\prime},s^{\prime}) \Gamma^{\mu} u(p,s) \chi^{\dag}_{s^{\prime}}
  \left( G_{E}(t) + \frac{i \vec{\sigma} \times \vec{q_{B}}}{2 m} G_{M}(t)  \right) \chi_{s},
  \ea
 where $\chi_{s}$ is the two-component of the Pauli spinor,
 $ G_{E}(t)$ and $G_{M}(t)$ are the Sachs form factors given by
 \ba
  G_{E}^{p/n}(t) = F_{1}^{p/n}(t)-\tau  F_{2}^{p/n}(t),
  \ea
  \ba
  G_{M}^{p/n}(t) = F_{1}^{p/n}(t) +  F_{2}^{p/n}(t),
  \ea
  where $\tau=t/(4 M^{2})$.
 Their three-dimensional Fourier transform  provides the electric charge density
 and the magnetic current density distribution \cite{Sachs62}.
  The form factors can be extracted from  experimental data on the elastic electron-nucleon
  scattering by the Rosenbluth method  \cite{Rosenbluth}  or
  from the polarization electron proton elastic scattering \cite{Akhiezer}.

 The radii of the particles are determined through the slope of the form factors at zero momentum transfer,
 and the squares of the Dirac radius $<r_{D}^{2}>$   and charge radius  $ <r_{E}^{2}>$  are determined as
   \ba
  <r_{D}^{2}> = -6 \frac{d F_{1}^{p/n}(t)}{dt}|_{t=0};  \  \ \ \ \
   <r_{E}^{2}> = -6 \frac{d F_{1}^{p/n}(t)}{dt}|_{t=0}+ \frac{3}{2}\frac{k_{p/n}}{m^{2}_{p/n}},
  \ea
%  Correspondingly for the squired charge radius is
%   \ba
%  <r_{D}^{2}> = -6 \frac{d F_{1}^{p/n}(t)}{dt}|_{t=0}+ \frac{3}{2}\frac{k_{p/n}}{m^{2}_{p/n}},
%  \ea
  There are some problems in the description of the form factors at small momentum transfer.
  To obtain a derivative of the form factors at $t=0$, we must have the analytic form of the form factors.
  However, such analytic forms are obtained by  fitting  experimental data in a wide
  region of momentum transfer. Hence, the form of the analytic function can be determined
  in most part by the data at non-small momentum transfer. As a result, such a form at very
  small $t$ can not satisfy the experimental data in the region of very small $t$.
  For example, the latest experimental data \cite{expPL17} show a large difference from the standard
  dipole form of the electromagnetic form factors at $0.01< Q^2 < 0.016$ GeV$^2$.
  If we take the standard dipole form of the electromagnetic form factors $\Lambda^4/(\Lambda^2-t)^2$
  with $\Lambda^2 =0.71 $ GeV$^2$,
  the analytic calculations give  $<r_{D}^{2}> = 0.76$ {\it fm} and  $<r_{E}^{2}> = 0.88$ {\it fm}.
  It coincides with recent data for $<r_{E}^{2}> = 0.875 (18)$ {\it fm} \cite{ep16}
   obtained from electronic hydrogen transition and   experimental data on electron scattering  \cite{ep16}  .
   Such parametrization takes into account only the experimental data
   for the charge proton form factor measured in the electron-proton scattering.
   There is a difference between the value of the proton radius  and the determination of  $<r_{E}^{2}>$
   in the muon-nucleon bound state, which gives $<r_{E}^{2}> = 0.84087(39)$ {\it fm} \cite{mu1,mu2}.
   This discrepancy, known as the "proton radius puzzle", gave birth to a  wide discussion and
    many approaches to explain this fact. But up to now
   the effect has no clear explanation.

 %        The study of the structure of the particle is one of the old and long stated
 %        problem in modern physics.
% In the modern physics main step was made by introducing the parton picture of the hadrons.
  The Parton distribution functions (PDF) are obtained  in the deep inelastic scattering
   using the recent data  at HERA and LHC.
   Beside this main point of the modern picture of the hadron structure, which depends
  only on the Bjorken longitudinal variable $x$,
  there were introduced a number of  other more complicated
  parton distributions which can depend on a number of  different variables.
%    The next step reveal some more
One of them is % the generelized structure function -
   the Generelized parton distribution function (GPDs) (which  depends on the Bjorken variable $\ x$,  momentum transfer $t$
  and skewnes parameter $\xi$) \cite{Mil94,Ji97,R97}.
% , transfer momentum distribution functions (TMDs) (which is depended over$x$, inner
%  momentum transfer $k$ and skuidenes parameters $\xi$) and many other.
  The conjunction between the momentum transfer and the impact parameter allow one to obtain
   the space parton distribution which has the probability conditions  \cite{Burk-15}.
   The connections between the deep-inelastic scattering, from which we can obtain the $x$-dependence of
   parton distributions, and the elastic electron-nucleon scattering, where the form factors of the nucleons
   are obtained, can be derived by  using
   the sum rules %introducing by
  \cite{Mil94,Ji97,R97}. % are connected
  The form factors, which are obtained in different reactions, can be calculated
  as   the  Mellin moments of GPDs. % with form factors of the different reactions.
 %   It is allow to connect the General parton distribution
%  functions with form factors of the hadrons which are related with different reactions %  including the inelastic
 % and elastic channels.
 Using the electromagnetic (calculated as the zero Mellin moment of GPDs) and gravitomagnetic form factors
 (calculated as the first moment of GPDs) in the hadron scattering amplitude, one can obtain a quantitative
 description of the hadron elastic scattering in a wide region of energy and transfer momenta.
 Many different forms
  of the $t$-dependence of GPDs were proposed.
   In the quark di-quark model \cite{Liuti1} the form of  GPDs
   consists of three parts - PDFs, function distribution and Regge-like function.
 In other works (see e.g. \cite{Kroll04}),
  the description of the $t$-dependence of  GPDs  was developed
  in a  more complicated representation using the polynomial forms with respect to $x$.

 \section{Momentum transfer dependence of GPDs }

To obtain the true form of the proton and neutron  form factors,
 it is important to have the true form of the momentum transfer dependence of GPDs.
Let us % modify the original Gaussian ansatz and
 choose  the $t$-dependence of  GPDs  in a simple form
%\ba
$ {\cal{H}}^{q} (x,t) \  = q(x) \   exp [  a_{+}  \
%    (1-x)^2/x^{m}
   f(x) \ t ],                                     $
%   \label{GPD0}
%\frac{(1-x)^2}{x^{m} } \ t ].
%\ea
  with $f(x)= (1-x)^{2}/x^{\beta}$
  %In
   \cite{ST-PRDGPD}.
The isotopic invariance can be used to relate the proton and neutron GPDs,
hence we have the same parameters for the proton and neutron GPDs.

%Add  ....................

   The complex analysis of the corresponding description of the electromagnetic form factors of the proton and neutron
    by  different  PDF sets  (24 cases) was carried out in \cite{GPD-PRD14}. These
   PDFs include the  leading order (LO), next to leading order (NLO) and next-next to leading order (NNLO)
   determination of the parton distribution functions.
    They used  different forms of the $x$ dependence of  PDFs. % eqs. (\ref{sq1})-(\ref{ex3}).
    We slightly complicated the form of GPDs  in comparison with the equation used in     \cite{ST-PRDGPD},  %       (\ref{GPD0}),
   but it is the simplest one as compared to other works (for example \cite{DK-13}).
%\ba
%{\cal{H}}^{u} (x,t) \  = q(x)^{u}  \   e^{2 a_{H}   f(x)_{u}  \ t };  \ \ \ % \\ \nonumber
%{\cal{H_d}}^{d} (x,t) \  = q(x)^{d}  \   e^{2 a_{H} f_{d}(x)  \ t };  \\
%%\label{t-GPDs-H}
%%\ea
%%\ba
%{\cal{E}}^{u} (x,t) \  = q(x)^{u} (1-x)^{\gamma_{u}} \   e^{2 a_{E}  \  f(x)_{u}  \ t }; \ \ \  % \\ \nonumber
%{\cal{E_d}}^{d} (x,t) \  = q(x)^{d}  (1-x)^{\gamma_{d}} \   e^{2 a_{E} f_{d}(x) \ t },
%\label{t-GPDs-E}
%\ea
% where
% $$ f_{u}(x) =  \frac{(1-x)^{2+\epsilon_{u}}}{(x_{0}+x)^{m}},$$ and $$f_{d}(x) = (1+\epsilon_{0}) (\frac{(1-x)^{1+\epsilon_{d}}}{(x_{0}+x)^{m}} % ).$$
%%  and  $q(x)^{u,d}_{fl}=q(x)^{u,d}_{nf} (1.-x)^{z_{1},z_{2}}$
%
%

 The hadron form factors will be obtained  by integration  over $x$ in the whole range of $x$ - $0 - 1$.
 Hence, the obtained  form  factors will be dependent on the $x$-dependence of the forms of PDF at the ends of the integration region.
 The  Collaborations determined the  PDF sets  from the inelastic processes only in  some region of $x$, which is only
 approximated to $x=0$ and $x=1$.
   Some  PDFs  have the polynomial form of $x$ with
     different power.  Some others have the exponential dependence of $x$.
  As a result, the behavior of  PDFs, when $x \rightarrow 0$ or $x \rightarrow 1$,  can  impact  the
    form of the calculated form factors.

    On the basis of our GPDs with, for example the PDFs
    ABM12 \cite{ABM12},    we calculated the hadron form factors
     by  numerical integration
   and then
    by fitting these integral results by the standard dipole form with some additional parameters
$$   F_{1}(t)  = (4m_p - \mu t)/(4m_p -  t ) \  \tilde{G}_{d}(t),   $$
% \label{Gt}
% \end{eqnarray}
  with
  $$
  \tilde{G}_{d}(t) = 1/(1 + q/a_{1}+q^{2}/a_{2}^2 +  q^3/a_{3}^3)^2 $$ which is slightly  different from
  the standard dipole form on two additional terms with small sizes of coefficients.
  The matter form factor %$A(t)$
\ba
 A(t)=  \int^{1}_{0} x \ dx
% \\	\nonumber
 [ q_{u}(x)e^{2 \alpha_{H} f(x)_{u} / t  } % \\	\nonumber
  + q_{d}(x)e^{ 2 \alpha_{H} f_{d}(x)  / t}  ] %$
\ea
 is fitted   by the simple dipole form  $  A(t)  =  \Lambda^4/(\Lambda^2 -t)^2 $.
        These form factors will be used in our model of the proton-proton and proton-antiproton elastic scattering
        and further in one of the vertices of the pion-nucleon elastic scattering.

%\section{Magnetic transition form factor  $G^{*}_{M (\gamma^{*} N \Delta ) } $    }

  To check the momentum dependence of the spin-dependent part of GPDs $ E_{u,d}(x,\xi=0,t) $,
   we can calculate the magnetic transition
  form factor \cite{sel-trff}, which is determined by the difference of $ E_{u}(x,\xi=0,t) $ and $ E_{d}(x,\xi=0,t) $.
 For the magnetic $N \rightarrow \Delta $ transition form factor $G^{*}_{M}(t)$,  in the large $N_{c}$  limit,
the relevant $GPD_{N\Delta}$ can be expressed in terms of the isovector GPD
    yielding the sum rules \cite{Guidal}
%\ba
%  G^{*}_{M} (t) = \frac{G^{*}_{M} (t=0)}{k_{v} } \int_{-1}^{1} dx ( E_{u}(x,\xi,t) - E_{d}(x,\xi,t) )
%\ea
%  where $k_{v}=k_{p} - k_{n} =3.70 $
% \ba
%  E_{u}(x,\xi=0,t) = d(x) e^{ [ 2 \alpha_{1} (\frac{ (1-x)^{p_{1} } }{ (x_{0} + x  )^{p_{2}} } ) ]}; \ \
%%  \ea
%%  \ba
%  E_{d}(x,\xi=0,t) = d(x) e^{ [ 2 \alpha_{1} (\frac{ (1-x)^{p_{1} k_{d} } }{ (x_{0} + x  )^{p_{2}} } + d x (1-x) t ) ]}.
%  \ea
%!   There are two different conventions - Jones-Scadron and Ash convention.
%  They are related as % The Jones-Scadron convention  is related to Ash conventional
%%\ba
%$G^{*}_{M, J-S}  (Q^{2}) = G^{*}_{M, Ash} \sqrt{1 +\frac{Q^{2} }{ (M+m)^{2} } } $.
%%\ea
%We will be present the data in the Ash convention.
%  The results of our calculations are presented in Fig.1 (a).
 The  experimental data  exist up to
   $-t =8 $ GeV$^2$ and our results  show a sufficiently good coincidence with experimental data.
   It is confirmed that the form of the momentum transfer dependence
    of  $E(x,\xi,t)$ determined in our model  is correct.
%\section{The Compton cross sections}
%%%%Fig 1
%\begin{figure}
%%\begin{center}
%%\mbox{\epsfysize=60mm\epsffile{GedGd.eps}}x
%%\includegraphics[width=.3\textwidth]{NDtff.ps} %{ffdm.ps}
%%\includegraphics[width=.3\textwidth]{Rvtkht.ps} %{ffdm.ps}
%\includegraphics[width=.45\textwidth]{Rvt.ps} %{ffdm.ps}
%\includegraphics[width=.45\textwidth]{Rat2kh.ps} %{ffdm.ps}
%\includegraphics[width=.3\textwidth]{Rat.ps} %{ffdm.ps}
%\includegraphics[width=.4\textwidth]{dsvtakh.ps} %{ffdm.ps}
%\vspace{0.5cm}
%\vspace{1.cm}
%\caption{a) [left] The transition magnetic  form factors  $G^{*}_{M, Ash}$
%(line- our calculations, points the experimental data \cite{G-data};
% b) [middle]
% the Compton form factors %(squares - show the model calculations \cite{DK-13} \\
%  $t^2 R_{V}(t)$ and  $t^2 R_{T}(t)$ , % b) [middle] $t^2 R_{T}(t)$  ,
%% b) [middle]  $t^2 R_{A}(t)$  .
% c) [right] $t^2 R_{A}(t)$
%  }
%\label{Fig_1}
%\end{figure}
%     Now let us calculate
 The moments of the GPDs with inverse power of $x$  gives us the
     Compton form factors.
      The results of our calculations of the Compton form factors
          %are shown in Fig. 1(a,b).
          coincide well with the existing experimental data.
%   Using the obtained form factors the reaction of the real Compton scattering
%    can be calculating.
%    The differential cross section for that reaction can be written as   \cite{DK-13} %%\cite{R98,DK-13}
% \ba
% \frac{d\sigma}{dt} =  \frac{\pi \alpha^{2}_{em}}{s^{2}} \frac{(s-u)^{2}}{-u s}
% [R_{V}^{2}(t) \ - \ \frac{t}{4 m^{2}} R^{2}_{T}(t) % \\ \nonumber
%   + \frac{t^{2}}{(s-u)^{2}} R^{2}_{A} (t)],
%   \label{RCS}
%ea
% where $R_{V}((t)$, $R_{T}(t)$, $R_{A}(t)$ are the form factors given by the $1/x$
% moments of the corresponding GPDs $H^{q}(x,t)$,  $E^{q}(x,t)$, $\tilde{H}^{q}(x,t)$ .
%    The last is related with the axial form factors.
%    As noted in \cite{DK-13}, this factorization,
% which bears some similarity to the handbag factorization of DVCS,
% is formulated in a symmetric frame where the skewness $\xi=0$.
  For  $H^{q}(x,t)$,  $E^{q}(x,t)$ we used the PDFs obtained from the
  works \cite{KKT12} with the parameters
  obtained in our fitting procedure of the description of the proton and neutron electromagnetic
   form factors   in \cite{GPD-PRD14}.
%    \ba
% R_{i}(t) =  \sum_{q} e^{2}_{q} \int_{0}^{1} \frac{dx}{x} {\cal{F}}j_{q}(x,\xi=0,t),
%\ea
% where ${\cal{F}}j_{q}$ are equal $H_{q}$, $E_{q}$ and $\tilde{H}_{q}$ and give the the form factors $R_{V}(t)$, $R_{T}(t)$, $R_{A}(t)$,
% respectively.  %  \ba
%
%  In the present work  for $\tilde{H}^{q}(x,t)$ we take $\Delta q^{e}$ in the form \cite{Khang-16}
% for  NNLO $Q_0=2$ GeV$^2 $
%\ba
% x \Delta_{q}(x,Q_{0}) = N_{q} \eta_{q} x^{a_{q} } (1-x)^{b_{q} } (1 + c_{q} x);
%\ea
% Assuming $SU(3)$ flavor symmetry of $\Delta \bar{q} $  the coefficient $N_{q}$ is determined as
%%\ba
%$ \frac{1}{N_{q}} = (1+c_{q} \frac{a_{q}}{1+a_{q}+b_{q}} ) \ B(a_{q},b_{q}+1)$,
%%\ea
% where $\ B(a_{q},b_{q}+1)$ is determined by
% $B(a,b)= \Gamma(a) \Gamma(b)/\Gamma(a+b) = \int_{0}^{1} t^{a-1} (1-t)^{b-1} dt $.
        %  Obviously,
%         $R_{V}(t)$ and $R_{T}(t)$ have a similar momentum transfer dependence but essentially differ in size.
%  On the contrary, the axial form factor $R_{A}$ has an essentially different $t$ dependence.
%         The results for the cross sections are presented in Fig.2 (a). Except the very %         large angles at low energy
%         the coincidence with the experimental data is sufficiently good.
%         The calculations of $R_{i}$  on the whole, correspond to the calculations %         \cite{DK-13}.

 \section{Electromagnetic and gravitomagnetic  form factors
            and the elastic nucleon-nucleon scattering }

        Both hadron  electromagnetic and gravitomagnetic form factors were used
        in the framework of the high energy generalized structure   (HEGS)
        model  of the elastic nucleon-nucleon scattering.
        This allowed us to build the model with a minimum number of fitting
        parameters \cite{HEGS0,HEGS1,NP-hP,NP-min}.

   The Born term of the elastic hadron amplitude can now be written as
  \begin{eqnarray}%\ba
 F_{h}^{Born}(s,t)=&&h_1 \ G^{2}(t) \ F_{a}(s,t) \ (1+r_1/\hat{s}^{0.5})
 %   \\ \nonumber
     +  h_{2} \  A^{2}(t) \ F_{b}(s,t) \     \\
     && \pm h_{odd} \  A^{2}(t)F_{b}(s,t)\ (1+r_2/\hat{s}^{0.5}),  \nonumber
    \label{FB}
% \nonumber
\end{eqnarray}
 where both (electromagnetic and gravitomagnetic) form factors are used.
 The parameters are determined in \cite{HEGS1}.
  The model is very simple from the viewpoint of the number of fitting parameters and functions.
  There are no any artificial functions or any cuts which bound the separate
  parts of the amplitude by some region of momentum transfer.
        In the framework of the model the description of  experimental data was obtained simultaniously
        at the large momentum transfer and in the Coulomb-hadron region in the energy region from $\sqrt{s}=9 $ GeV
        up to LHC energies with taking into account the Coulomb-nuclear phase \cite{selmp1,Selphase}.
        The model gives a very good quantitative description of  recent
        experimental data at $\sqrt{s}=13$ TeV \cite{Sel-PL19}.

\section{Nucleon gravitomagnetic  radii }

 % Using the obtained GPDs of the nucleon
 A good description of the various form factors and the elastic scattering of the hadrons
 gives a good support to  our determination of the momentum transfer dependence of GPDs.
 Based on this determination of GPDs,
   let us calculate the gravitomagnetic radius of the nucleon using the integral representation of the form factor
  and make the numerical differentiation over $t$ at $t \rightarrow 0$. This method allows us
  to obtain a concrete form of the form factor by fitting the result of the integration of
   the GPDs over $x$.
   As a result, the gravitomagnetic radius is determined as
   \ba
<{r_{A}}^2> =  -\frac{6}{A(0)} \frac{dA(t)}{dt}|_{t=0};
\label{Rp}
\ea
hence the numerical derivative will be
   \ba
<{r_{A}}^2> =  -\frac{6}{A(0)} \frac{A(t_{1})-A(t_{1}+\Delta t}{\Delta t}|;
\label{Rp}
\ea
   where $A(t) = \int_{0}^{1} x \ 3 (q_{u}(x) + q_{d}(x)) \ e^{-\alpha \ t f(x)} dx$

     The GPDs will be taken with   various  forms of the PDFs obtained by
    different Collaborations (see our paper \cite{GPD-PRD14}).
   To compare the $t$ dependence of the starting point of the differentiation of the results, we take 3 variants
  $$ 1) -t_{1} =1. 10^{-3} \ \ \ {\rm and} \ \ \ \Delta t_{1} =1. 10^{-3};$$
  $$ 2)-t_{1} =1. 10^{-2}  \ \ \ {\rm and} \ \ \   \Delta t_{1} =1. 10^{-3}; $$
  $$  3)-t_{1} =4. 10^{-2} \ \ \  {\rm and}  \ \ \  \Delta t_{1} =1. 10^{-3};  $$

\begin{table}
%\begin{table}
 \caption{The  gravitation radius of the nucleon }
\label{Table-1}
%\vspace{.1cm}
\begin{center}
\begin{tabular}{|c||c|c|c||c|c|c|} \hline
%             &               &              &  &    &          \\
            &    \multicolumn{3}{c||} {} &   \multicolumn{3}{c|} {}      \\
  & \multicolumn{3}{c||} {$ (<{r_{A}}^2)>^{1/2} basic \ variant $ } & \multicolumn{3}{c|} { $ (<{r_{A}}^2)>^{1/2} basic + 4 \ parameters$ }  \\
  &     \multicolumn{3}{c||} { }  &      \multicolumn{3}{c|} { }    \\ \hline
 Model  & $-t_1$ & $ -t_1$ & $-t_1$   & $-t_1$ & $-t_1$ & $-t_1$  \\ %  &  Order(Q)        \\
  of PDFs      & $ 10^{-5}$ & $ 10^{-3} $& $4 \ 10^{-2}$   & 0.001& 0.001 & 0.02          \\
   & $ \Delta t=10^{-5}$ & \multicolumn{2}{c||} { $ \Delta t=10^{-3}$ } &$ \Delta t=10^{-5}$   & \multicolumn{2}{c|} { $ \Delta t=10^{-3} $ }  \\
     & GeV$^2$ &  GeV$^2$  &  GeV$^2$   &  GeV$^2$  &  GeV$^2$  & GeV$^2$          \\
%N & Model PDFs  & $p_1$ & $p_2$ &    &  \\
%              &               &       & & & &                    \\ \hline
%              &               &            &      & & & &       \\
            &              &                &      & & &      \\
 \hline
 ABKM09   & 0.525 & 0.525 & 0.505 & 0.536 & 0.535 &0.522  [fm]  \\ %& NNLO (3.)      \\
 JR08a    & 0.546 & 0.545 & 0.547 & 0.566 & 0.565 & 0.543 [fm]  \\ %& NNLO (0.55)    \\
 JR08b    & 0.526 & 0.525 & 0.507 & 0.547 & 0.547 & 0.527 [fm]  \\ %& NNLO (2.)      \\
 ABM12    & 0.524 & 0.524 & 0.505 & 0.543 & 0.543 & 0.523 [fm]  \\ %& NNLO (0.9)     \\
 KKT12a   & 0.525 & 0.524 & 0.504 & 0.525 & 0.525 & 0.507 [fm]  \\ %& NLO (4.)       \\
 KKT12b   & 0.519 & 0.519 & 0.521 & 0.498 & 0.518 & 0.501 [fm]  \\ %& NLO (4.)       \\
 GJR07d   & 0.523 & 0.523 & 0.505 & 0.531 & 0.531 & 0.515 [fm]  \\ %& LO (0.3)       \\
 GJR07b   & 0.532 & 0.531 & 0.512 & 0.546 & 0.544 & 0.526 [fm]  \\ %& NLO (0.3)      \\
 GJR07a   & 0.506 & 0.506 & 0.498 & 0.521 & 0.520 & 0.503 [fm]  \\ %& LO (0.3)       \\
 GJR07c   & 0.510 & 0.510 & 0.501 & 0.519 & 0.519 & 0.502 [fm]  \\ %& NLO (0.3)      \\
 % 6a& MRST02  & \cite{MRST02}& Eq. (\ref{sq2a})& NLO (1.)        & & & &    \\
 % 6b& MRST01  & \cite{MRST01}& Eq. (\ref{sq2a})& NLO (1.)        & & & &        \\
 % 7a& GP08a    & \cite{GP08}  & Eq. (\ref{sq2b})& NLO (0.5)      \\
 % 7b& GP08b    & \cite{GP08}  & Eq. (\ref{sq2b})& NNLO (1.5)    \\
 % 7c& GP08c    & \cite{GP08}  & Eq. (\ref{sq2b})& NLO (2.)   \\
 % 7d& GP08d    & \cite{GP08}  & Eq. (\ref{sq2b})& NNLO (0.5)      \\
 % 8a& MRST09  & \cite{MRST09}& Eq. (\ref{sq2a})&  LO   (1.)       \\
 % 8b& MRST09  & \cite{MRST09}& Eq. (\ref{sq2a})&  NLO  (1.)        \\
 % 8c& MRST09  & \cite{MRST09}& Eq. (\ref{sq2a})& NNLO  (1.)         \\
 % 9 & MRST02P  & \cite{CTEQ6M}& Eq. (\ref{ex1})& NLO (1.3)       \\
% 0a& CJ12amin    & \cite{CJ12}  & Eq. (\ref{sq2a})& NLO(1.7)        \\
% 0b& CJ12am    & \cite{CJ12}  & Eq. (\ref{sq2a})& NLO(1.7)        \\
% 0c& CJ12bmid    & \cite{CJ12}  & Eq. (\ref{sq2a})& NLO(1.7)       \\
% 0c& CJ12cmax    & \cite{CJ12}  & Eq. (\ref{sq2a})& NLO(1.7)       \\
% 1 & MRSTR4  & \cite{MRST02}& Eq. (\ref{sq2a})&    NLO (1.3)        \\
            &              &                &      & & &      \\
 \hline
\end{tabular}
\end{center}
%\end{table}
  \end{table}

   To examine the  dependence of the results on our model parametrization of GPDs,
     we  made the calculation for the simplest basic variant with minimum parameters
    and a more complicated variant with  additional 4 parameters.
   It was shown in \cite{GPD-PRD14} that the final result of the fitting of GPDs
   on the basis of practically all experimental data on the electromagnetic form factors of the proton and
   neutron, weakly depended on  adding the supplementary fitting parameters but some others
   had a heavy dependence. We put the GPDs of the first class     in the upper rows of  Table 1.
    It is clear from the comparison of the variants 1) and 2) (see Table 1)
     that there is no  difference between
    the results obtained at $-t=10^{-5}$ GeV$^2$ and $-t=10^{-3}$ GeV$^2$.
    However, already at $-t=4. \ 10^{-2}$ GeV$^2$ the results have  significant differences.
    Hence, if we make the fitting procedure to obtain the analytic form of the form factors,
    the region of a very small momentum transfer has to be leading.

%   The comparizon of the "basic"  with "basic + 4 parameters" variants
%   shows that for PDFs of the various Collaborations leads to the different
%   results.
   Note that the "basic"  and the "basic + 4 parameters" variants can lead
   to the same different results for the PDFs of  different Collaborations.
For example, the PDFs of (KKT12a \cite{KKT12}) give the radius which is practically independent of
   additional free parameters. Hence, this result can be considered as most stable and probable.
   Note that this size of the radius practically coincides with the arithmetic means
   of all radii (in the first column), which is equal to $\bar{(<{r_{A}}^2)>^{1/2}} =0.524$ {\it fm}.

\section{Nucleon electromagnetic  radii }

 Now let us calculate the Dirac radius of the proton    $<r_{D}^{2}>$
 and the charge radius of the proton   $<r_{E}^{2}>$
  using our % form and
  momentum transfer dependence of GPDs.
     We used the same procedure as for our calculations of the matter radius.
  As a  result, the Dirac radius is determined from the zero Mellin moment of GPDs
   \ba
<{r_{D}}^2> =  -\frac{6}{F(0)} \frac{dF(t)}{dt}|_{t=0};
\label{Rp}
\ea
hence the  differentiation %  derivative
 will be
   \ba
<{r_{D}}^2> =  -\frac{6}{F(0)} \frac{F(t_{1})-F(t_{1}+\Delta t)}{\Delta t}|_{t \rightarrow 0};
\label{Rp}
\ea
    where
    \ba
    F(t) = \int_{0}^{1} ( e_{u} \ q_{u}(x) \ + \ e_{d} \  q_{d}(x)) \ e^{-\alpha \ t f(x)} dx.
    \ea
  Now   we consider only the first variant with $t_{1}= 10^{-5}$ GeV$^2$ and    $\Delta t= 10^{-5}$ GeV$^2$.

     The results are presented in Table 2.
  Again, we see that a variant of PDFs gives the result which weekly depends on the
  number of  free parameters. If the arithmetic mean  of $(<{r_{e}}^2>^{1/2})$ is calculated,
  we find that the charge radius  $(\bar{<{r_{e}}^2>^{1/2}}) = 0.850$ {\it fm} for the basic variant
  and  $(\bar{(<{r_{e}}^2>^{1/2}} = 0.860)$ {\it fm}  for the variant with additional four fitting parameters.
  These values are between the sizes of the charge proton radii  determined in the electron and muon
  experiments. In some sense, this  removes the discrepancy between these two methods.
  But, of course,  our errors, especially which come from the $x$-dependence of different PDFs
  at $x \rightarrow 0$ and  $x \rightarrow 1$,
  are large, and this requires further deep analysis.

\begin{table}
%\begin{table}
 \caption{The  electromagnetic radii of the nucleon }
\label{Table-1}
%\vspace{.1cm}
\begin{center}
\begin{tabular}{|c||c|c|c|c||c|c||c|} \hline
%             &               &              &  &    &          \\
%  & \multicolumn{3}{c} { $ basic \ variant $ }& \multicolumn{3}{c} { $ basic + 4 \ parameters$ }& \\ \hline
% N & Model  & Reference&  $-t_1$ & $ -t_2$ & $-t_3$   & $-t_1$ &   Order(Q)        \\
            &  & & \multicolumn{2}{c|} { } &    \multicolumn{2}{c||} { }   &      \\
  N & Model  & Refer.& \multicolumn{2}{c|} {$<r_{D}>$ }&\multicolumn{2}{c||}{ $<r_{G}>$  } &   Order(Q)        \\
 &  &  & $ basic $ & $+4_{par.}$ & $ basic$ & $+ 4_{par.}$ &  \\
%        & $ 10^{-5}$ & $ 10^{-3} $& $4 \ 10^{-2}$   & 0.001& 0.001 & 0.02 &         \\
%   & $ \Delta t=10^{-5}$ & \multicolumn{2}{c} { $ \Delta t=10^{-3}$ } &$ \Delta t=10^{-5}$   & \multicolumn{2}{c} { $ \Delta t=10^{-3} $ } & \\
%     & GeV$^2$ &  GeV$^2$  &  GeV$^2$   &  GeV$^2$  &  GeV$^2$  & GeV$^2$ &         \\\hline
%N & Model PDFs  & $p_1$ & $p_2$ &    &  \\
%              &               &       & & & &                    \\ \hline
%              &               &            &      & & & &       \\
            &              &                &      & & & &      \\
 \hline
 1  & ABKM09  & \cite{ABKM09} & 0.787 & 0.787 & 0.859 & 0.859  [fm]&   NNLO (3.)      \\
 2a & JR08a   & \cite{JR08}   & 0.808 & 0.793 & 0.879 & 0.865  [fm]&   NNLO (0.55)     \\
 2b & JR08b   & \cite{JR08}   & 0.789 & 0.793 & 0.861 & 0.865  [fm]&   NNLO (2.)       \\
 3  & ABM12   & \cite{ABM12}  & 0.778 & 0.787 & 0.851 & 0.859  [fm]&   NNLO (0.9)     \\
 4a & KKT12a  & \cite{KKT12}  & 0.788 & 0.789 & 0.860 & 0.861  [fm]&   NLO (4.)         \\
 4b & KKT12b  & \cite{KKT12}  & 0.780 & 0.796 & 0.853 & 0.867  [fm]&   NLO (4.)        \\
% 5a & GJR07d  & \cite{GJR07}  & 0.523 & 0.523 & 0.505 & 0.531  [fm]&   LO (0.3)      \\
 6a & GJR07d  & \cite{GJR07}  & 0.735 & 0.781 & 0.811 & 0.853  [fm]&   NLO (0.3)      \\
 5b & GJR07b  & \cite{GJR07}  & 0.788 & 0.781 & 0.860 & 0.853  [fm]&   NLO (0.3)     \\
 5c & GJR07a  & \cite{GJR07}  & 0.766 & 0.775 & 0.840 & 0.848  [fm]&   LO (0.3)      \\
 5d & GJR07c  & \cite{GJR07}  & 0.748 & 0.775 & 0.824 & 0.848  [fm]&   NLO (0.3)      \\
% 6b  % 6a& MRST02  & \cite{MRST02}& Eq. (\ref{sq2a})& NLO (1.)        & & & &    \\
% 7a  % 6b& MRST01  & \cite{MRST01}& Eq. (\ref{sq2a})& NLO (1.)        & & & &        \\
% 7b  % 7a& GP08a    & \cite{GP08}  & Eq. (\ref{sq2b})& NLO (0.5)      \\
% 7c  % 7b& GP08b    & \cite{GP08}  & Eq. (\ref{sq2b})& NNLO (1.5)    \\
% 7d  % 7c& GP08c    & \cite{GP08}  & Eq. (\ref{sq2b})& NLO (2.)   \\
 % 7d& GP08d    & \cite{GP08}  & Eq. (\ref{sq2b})& NNLO (0.5)      \\
 % 8a& MRST09  & \cite{MRST09}& Eq. (\ref{sq2a})&  LO   (1.)       \\
 % 8b& MRST09  & \cite{MRST09}& Eq. (\ref{sq2a})&  NLO  (1.)        \\
 % 8c& MRST09  & \cite{MRST09}& Eq. (\ref{sq2a})& NNLO  (1.)         \\
 % 9 & MRST02P  & \cite{CTEQ6M}& Eq. (\ref{ex1})& NLO (1.3)       \\
% 0a& CJ12amin    & \cite{CJ12}  & Eq. (\ref{sq2a})& NLO(1.7)        \\
% 0b& CJ12am    & \cite{CJ12}  & Eq. (\ref{sq2a})& NLO(1.7)        \\
% 0c& CJ12bmid    & \cite{CJ12}  & Eq. (\ref{sq2a})& NLO(1.7)       \\
% 0c& CJ12cmax    & \cite{CJ12}  & Eq. (\ref{sq2a})& NLO(1.7)       \\
% 1 & MRSTR4  & \cite{MRST02}& Eq. (\ref{sq2a})&    NLO (1.3)        \\
            &              &                &      & & & &      \\
 \hline
\end{tabular}
\end{center}
%\end{table}
  \end{table}

%\ba
%{\cal{R}}^2 =  -\frac{6}{A(0)} \frac{dA(t)}{dt}|_{t=0};
%\label{Rp}
%\ea

\section{Conclusion}

  The structure of the hadron represented by the Generelazed parton distribution functions
  is  the main part in the amplitudes of  different reactions.
  The Mellin moments of GPDs allow us to calculate  different form factors
  from the same analytic functions.
  The parameters of the phenomenological form
  of GPDs can be obtained from the analysis of the experimental data of the
  proton and neutron electromagnetic form factors simultaneously.
  Our determination  of the momentum transfer dependence of GPDs of hadrons
  allows us to obtain  good quantitative descriptions of
   different form factors, including
    the Compton, electromagnetic,
     transition and gravitomagnetic form factors  simultaneously.
  Hence, our calculations
  of the nucleons radii are based on the use of the maximum number
  of  experimental data.

  Our method of the calculation of the radius  from the numerical derivation
  of the integral of GPDs over variable $x$ does not require any fitting
  of  separate hadron form factors.
   Most significant uncertainty in the determination
  of the nucleon radii comes from the   indeterminacy of the form of the parton
  distribution function which has the phenomenological origin with the parameters
  determined from the analysis of  different deep inelastic reactions.
  Our calculations show  the low values of the charge radius of the proton compared
  to the standard determination of the dipole form of the proton form factor.
  In some sense,  this removes the contradictions with the size of the charge
  radius of the proton obtained in the muon experiment.
  Of course, the uncertainty which comes from  different forms of the PDFs
  is large now, and this requires  further analysis.

\vspace{1.5cm}
{\bf Acknowledgments}

%\section{Acknowledgments}
 {\it The authors would like to thank J.-R. Cudell and O.V. Teryaev
   for fruitful   discussion of some questions   considered in the paper.}

%\section*{References}


\begin{thebibliography}{9}
%%\bibitem{iopartnum} IOP Publishing is to grateful Mark A Caprio, Center for Theoretical Physics, %Yale University, for permission to include the {\tt %iopart-num} \BibTeX package (version 2.0, %December 21, 2006) with  this documentation. Updates and new releases of {\tt iopart-num} can be %found on %\verb"www.ctan.org" (CTAN).
\bibitem{Lomon} E.L. Lomon, S. Pacetty   Phys.Rev. D {\bf 86}, 039901  (2012).
%\bibitem{Foldy} L.L. Foldy, Phys.Rev. {\bf 87},  688  (1952).

      \bibitem{QA12} % G.D. Gates {\it et al.}, Phys.Rev.Lett. {\bf 106} 252003 (2011);
       I.A. Qattan and J. Arrington, Phys.Rev. C {\bf 86}, 065210 ( 2012).


%4
\bibitem{Ernst60} F.J. Ernst, R.G. Sachs, and K.C. Wali Phys. Rev. {\bf 119},  1105  (1960).
%5
\bibitem{Sachs62}R.G. Sachs, Phys.Rev. {\bf 126},  2256  (1962).
\bibitem{Kelly-02} J.J. Kelly, Phys.Rev. C {\bf 66}, 065203 (2002).

\bibitem{Rosenbluth} M.N. Rosenbluth, Phys.Rev. {\bf 79} 615 (1950).
\bibitem{Akhiezer} A.I. Akhiezer and M.P. Rekalo Sov.J.Pat.Nucl. {\bf 3}  277  (1974).

  \bibitem{expPL17} M. Mihovilovic et al. , Phys.Lett., {bf B 771} 194 (2017).

  % \bibitem{PDG16}  Particle Data Gropp (1916).
    \bibitem{ep16} P.J. Mohr, D.B. Newell, and B.N. Taylor,
    Rev.Mod.Phys. {\bf 88}, 035009 (2016).

    \bibitem{mu1}  R. Pohl et al., Nature {\bf 466} 417 (2010)
    \bibitem{mu2}  M. Mihovilovic et al., Phys.Lett,, {bf 771} 194 (2017).



%%13
%\bibitem{Arnold} R.G. Arnold, C.E. Carlson, and F. Gross,
%             Phys.Rev. C {\bf 23},   363  (1981).
%%10
%\bibitem{Guichon} P.A.M. Guichon, M. Vanderhaeghen,
%  Phys.Rev,Lett. {\bf 91}, 142303-1   (2003).
%  P.G. Blunden, W. Melnitchouk, and A. Tjon,
%  Phys.Rev,Lett. {\bf 91}, 142304 (2003);
%   Y.-C. Chen, A. Afanasev, S. Brodsky, C. Carlson, and  M. Vanderhaeghen,
%    Phys.Rev.Lett. {\bf 93}, 122301 (2004);
%  M.P. Recalo, E. Tomasi-Gustafsson, Eur.Phys.J. A {\bf 22} 119 (2004);
%  S. Dubnichka, E. Kuraev, M. Secansky, A. Vinnikov, arXiv:hep-ph/0507242.
%
%
%
%









%\bibitem{Rev-LHC} R. Fiore, L. Jenkovszky, R. Orava, E. Predazzi, A. Prokudin, O. Selyugin,
%                    Mod.Phys., A24  (2009) 2551. %-2559
%
%
%  \bibitem{Meis-08} S. Meissner, A. Metz, M. Schlegel, and K. Goeke,
%    JHEP, {\bf 0908} (2009) 056.
%
%%  \bibitem{Lorce-13} C. Lorce and B. Pasquini,
%          JHEP, {\bf 1309} (2013) 138.

  \bibitem{Burk-15} M. Burkardt and B. Pasquini,
        EPJA Special Issue on "3D Structure of the Nucleon"; EPJ,


 % \bibitem{DMuller94} D. Muller, D. Robaschik, B. Geyer, F.M. Dittes and J. Horejsi, Fortsch. Phys. %{\bf 42}, (1994) 101;
\bibitem{Mil94} D. Muller {\it et al.}, Fortsch. Phys. {\bf 42}, (1994) 101;
%1x
 \bibitem{Ji97} X.D. Ji, Phys. Lett. {\bf 78} , (1997) 610; Phys. Rev D {\bf 55} (1997) 7114;

   \bibitem{R97}   Radyushkin, A.V.,  Phys. Rev. D {\bf 56},  5524 (1997).
% \bibitem{R97}  Radyushkin, A.V.,  Phys. Rev. D {\bf 56},  5524 (1997)  .
%\bibitem{Stol01} P. Stoler,  Phys.Rev., D {\bf 65} (2002) 053013.
%  % hep-ph/0108257.
%\bibitem{Stol02} P. Stoler,  Phys.Rev.  Lett.,  {\bf 91} (2003) 172303.
 % hep-ph/0210184.

%   \bibitem{Burk-15} M. Burkardt and B. Pasquini,
%        EPJA Special Issue on "3D Structure of the Nucleon"; EPJ,




\bibitem{Liuti1} G.R. Goldstein, J.O. Hernandez, S. Liuti,  Phys.Rev. {\bf D84} 034007 (2011).
% arXiv:1206.1876 v3.

%\bibitem{Liuti2}J.O. Gonsales-Hernandes  {\it et al.}, arXiv:1206.1876 v3.

%
 \bibitem{Kroll04}  M.Diehl {\it et al.},  Eur.Phys. J. C  {\bf 39} (2005) 1.
%% arXiv [hep-ph/0408173].

%\bibitem{Yuan03} F. Yuan, Phys. Rev. D, {\bf 69}, 051501(R) (2004) .
%% arXiv [hep-ph/0311288].

\bibitem{ST-PRDGPD}  O. Selyugin, O. Teryaev, Phys. Rev.
  {\bf D 79} 033003 (2009); % arXiv [0901.1786]:


% \bibitem{MRST02}  A.D. Martin  {\it et al.}, Phys. Lett. B  {\bf 531} (2002) 216.

 %35
 \bibitem{GPD-PRD14}  O.V. Selyugin,
  % Models of parton distributions and the description of form factors of nucleon
       Phys. Rev. {\bf D 89} 093007 (2014) . % arXiv [1404.2702]:



%    \bibitem{DK-99}    M. Diehl, T. Feldmann, R. Jakob and P. Kroll,
%          Eur.Phys.J. {\bf C8} 409 (1999). % [hep-ph/0408173]
%36
      \bibitem{DK-13}M. Diehl and P. Kroll,
          Eur.Phys.J. {\bf C73} 2397 (2013).

\bibitem{ABM12}  S. Alekhin, J. Blu"mlein, and S. Moch, Phys.Rev. D86
, 054009 (2012). % arXiv:1202.2281.


\bibitem{sel-trff}   O.V. Selyugin, in Procceding of XXVII Int. workshop "Spin in high energy physics",
  Dubna (2017); J.Phys.Conf.Ser. 938 (2017).


\bibitem{Guidal} M. Guidal, M. V. Polyakov, A. V. Radyushkin, M. Vanderhaeghen,
               PhysRevD. {\bf D72}  054013 (2004). % 054013.



%\bibitem{R98}   Radyushkin, A.V.,  Phys. Rev. D {\bf 58},  114008 (1998).

% ??         \bibitem{Kh12}  H. Khanpour {\it et al.}, arXiv:1205.5194
             \bibitem{KKT12}  H. Khanpour {\it et al.}, arXiv:1205.5194

%\bibitem{80-Flo-09}  D. de Florian, R. Sassot, M. Strtmann, W. Vogelsang,
  % Models of parton distributions and the description of form factors of nucleon
%       Phys. Rev. {\bf D 80} 034030 (2009) . % arXiv [1404.2702]:

%\bibitem{Khang-16}     F.Taghavi-Shahri, H. Khanpour .. 1603.03157



%24
%   \bibitem{HEGS-JEP12} O.V.~Selyugin,
      \bibitem{HEGS0} O.V.~Selyugin,
      Eur.Phys.J. {\bf C72}, 2073 (2012).


%35
 \bibitem{HEGS1}  O.V. Selyugin,
  % Models of parton distributions and the description of form factors of nucleon
       Phys. Rev. {\bf D 91 } 113003 (2015). % arXiv []:

%%25
      \bibitem{NP-hP}  O.~V.~Selyugin,
%     "Hard and soft pomerons in the elastic nucleon scattering", \\
         Nucl.Phys. A {\bf 903} 54 (2013). % 54-64.
         % arXiv:1205.5867

      \bibitem{NP-min}       O.V. Selyugin,    Nucl.Phys. {\bf A 959 } 116  (2017).


% \bibitem{Sel-Df16}     O.V. Selyugin,  "Diffraction (2016)


%   %27
\bibitem{selmp1} O.V. Selyugin, Mod. Phys. Lett. A{\bf 9} 1207 (1994).
   %28
%\bibitem{selmp2} O.V. Selyugin, Mod. Phys. Lett. A{\bf 14}, 223 (1999).
   %29
\bibitem{Selphase}
   O.\,V.~Selyugin,
  %``Coulomb Hadron Phase Factor And Spin Phenomena In A Wide Region Of Transfer
  %Momenta,''
  Phys.\ Rev.\  D {\bf 60} (1999) 074028
  %%CITATION = PHRVA,D60,074028;%%

\bibitem{Sel-PL19}      O.~V.~Selyugin,     Phys.Lett. {\bf B 797 } 1134870 (2019).



% \bibitem{Unit-PRD} 	J.-R. Cudell, E. Predazzi, O. V. Selyugin,
  %"New analytic unitarisation schemes", \\
%       Phys.Rev. {\bf D 79  }, 034033  (2009). % [arXiv:0812.0735[hep-ph]].

%\bibitem{G-data} F. Hagelstein, arxiv: 1710.00874.

%\bibitem{105-Dan07} A. Danagoulian, et. al. (Jefferson Lab Hall A Collaboration),
% Phys.Rev.Lett., {\bf 98} 152001 (2007).

% \bibitem{HEGS0} O.V.~Selyugin, Eur.Phys.J. C72 (2012) 2073.
 %9
% \bibitem{HEGS1}  O.V. Selyugin,
%  % Models of parton distributions and the description of form factors of nucleon
%       {\it Phys. Rev. D}, {\bf 91}, 11303 (2015). % V.91. P.113003. % arXiv []:
%\bibitem{HEGS1}  O.V. Selyugin,
  % Models of parton distributions and the description of form factors of nucleon
%    Phys. Rev. D 91 (2015) 11303. % V.91. P.113003. % arXiv []:



%\bibitem{Kuraev-SF} M.V. Galynskii, E.A. Kuraev, Phys.Rev. D, {\bf 89} (2014) 054005.

% \bibitem{W-Kur}  O.V. Selyugin, Particel Nucleii Letters

%37
   \bibitem{JR08} M. Gluck,
   Phys.Rev., {\bf D79 }, 074023 (2009).
   % [0810.4274]
%38
% \bibitem{GP08} C. Pisano,
% Phys.Rev., {\bf D77 }, 074002 (2008;
%   Erratum-ibid  {\bf D78} 019902(E) (2008).
  %[0812.3250]






% \bibitem{M09} Martin-09  0901.0002


%          \bibitem{KKT12}  H. Khanpour {\it et al.}, arXiv:1205.5194

%  \bibitem{CTEQ6M} J. Pumplin, {\it et al.}, JHEP 0207:012 (2002).

     \bibitem{ABKM09}S. Alekhin {\it et al.}, Phys.Rev. {\bf D81}
, 014032 (2010). % arXiv:0908.2766. %    \bibitem{ABM11}

% \bibitem{ABM12}  S. Alekhin, J. Blu"mlein, and S. Moch, Phys.Rev. {\bf D86}
%, 054009 (2012). % arXiv:1202.2281.

%43
\bibitem{GJR07} M. Gluck, P. James-Delgado, E. Reya,
                     Eur.Phys.J., {\bf C53} 355 (2008).


%0709.0614 - Gluck



%\bibitem{BurkT} M. Burkhardt, [hep-ph]/0509316.

%\bibitem{SopperT} D.E. Soper, Phys.Rev., D {\bf 15} (1977)1141.

%\bibitem{RayT} H. Dahiya, A. Mukherjee, S. Ray, [hep-ph]/0705.3580.

%\bibitem{Vanderh} C.E. Carlson and M. Vanderhaeghen,
%             Phys.Rev.Lett., {\bf 100} (2008) 032004.

%\bibitem{BurkZ} M. Burkhardt, [hep-ph]/0709.2966v2(October, 2007).

%\bibitem{MillerZ} G. A. Miller,
%  Phys.Rev.Lett., {\bf 99} (2007) 112001.



% \bibitem{Mez16} C. Mezrag, H. Moutarge, J. Rodriguez-Quintero, F. Sabatie,
%    ???Phys.Rev. {\bf D 89}, 074506 (2014)???.


% \bibitem{ETM}  ETM Collaboration, R. Frezzotti, V. LubicZ, S. Simula, Phys.Rev. {\bf D 79}, 074506 (2009).
% [hep-ph]/0705.2409v2(May, 2007); [nucl-th]/0802.2563v2(20 Feb. 2008).
%60
%   \bibitem{Mez14} C. Mezrag, H. Moutarge, J. Rodriguez-Quintero, F. Sabatie,
%    ???Phys.Rev. {\bf D 89}, 074506 (2014)???.

%   \bibitem{Dane14} A. Dane




%\bibitem{Arrington-ron} Kelly J.J., Phys. Rev. C, {\bf  66}, 065203 (2002)


%23
%\bibitem{SCP-EPJ08}
%    O.V. Selyugin, J.-R. Cudell, E. Predazzi,
 %"Analytic properties of different unitarization schemes",
%  Eur.Phys.J.ST,  162 (2008) 37.

%33
%\bibitem{Disser-data} Durham~HepData~Project, M.R. Whalley,
%\bibitem{data-Sp}
%http://durpdg.dur.ac.uk/hepdata/reac.html.
%34
%\bibitem{Land-Bron} K.R. Schubert, In Landolt-Bronstein, New Series, v. 1/9a, (1979).


%35
%\bibitem{TOTEM-8nexp} The TOTEM Collaboration (G. Antchev et al.)
       %  {\it Nucl. Phys. B}, {\bf 899}, 527 (2015).
%        Nucl. Phys. B  899 (2015) 527.
         % Tabl-data

 %\bibitem{TOTEM-11}
% \bibitem{T7a}  G. Antchev et al. (TOTEM Coll.)
 %                  {\it Eur.Phys.Lett.}, {\bf 96} 21002 (2011).
%  Eur.Phys.Lett., 96 (2011) 21002.
               % arhiv:1110.1395   7TeV t=0.02


  %\bibitem{TOTEM-1395}
%   \bibitem{T7b} G. Antchev et all. (TOTEM Coll.)
     %      {\it Eur.Phys.Lett.}, {\bf 101} 21002 (2013).
%    Eur.Phys.Lett.  101 (2013) 21002.
               % arhiv:






%\bibitem{T8a} G. Antchev et all. (TOTEM Coll.), CERN-PH-EP-2015-325.
           %TOTEM-15


%    \bibitem{Diff16} F. Nemes (TOTEM Collaboration), talk in the  Workshop
%          on Diffraction in High-Energy Physics, Aciriale (Italy), September 3-8 (2016),

%    \bibitem{Blois17} M. Deile  (TOTEM Collaboration), talk in the  Workshop
%          on Diffraction in High-Energy Physics, Praha (Czech. Resp.), June 26-30 (2017).


\end{thebibliography}
\end{document}